# A light-driven three-dimensional plasmonic nanosystem that translates molecular motion into reversible chiroptical function

Anton Kuzyk[1,*], Yangyang Yang[2,3,†], Xiaoyang Duan[1,4], Simon Stoll[1], Alexander O. Govorov[5], Hiroshi Sugiyama[2,3], Masayuki Endo[2], and Na Liu[1,4,*]

[1]*Max Planck Institute for Intelligent Systems, Heisenbergstrasse 3, D-70569 Stuttgart, Germany*
[2]*Institute for Integrated Cell-Material Sciences (WPI-iCeMS), Kyoto University, Yoshida-ushinomiyacho, Sakyo-ku, Kyoto 606-8501, Japan*
[3]*Department of Chemistry, Graduate School of Science, Kyoto University, Kitashirakawa-oiwakecho, Sakyo-ku, Kyoto 606-8502, Japan*
[4]*Kirchhoff Institute for Physics, University of Heidelberg, Im Neuenheimer Feld 227, D-69120 Heidelberg, Germany*
[5]*Department of Physics and Astronomy, Ohio University, Athens, Ohio 45701, USA*[†]*Present address: Shanghai Key Laboratory of Chemical Biology, School of Pharmacy, East China University of Science and Technology, 130 Meilong Road, Shanghai 200237, P. R. China*

[*]*To whom the correspondence should be addressed. E-mail: kuzyk@is.mpg.de and laura.liu@is.mpg.de*

**Abstract**

**Nature has developed striking light-powered proteins, such as bacteriorhodopsin, which can convert light energy into conformational changes for biological functions. Such natural machines are a great source of inspiration for creation of their synthetic analogs. However, synthetic molecular machines typically operate at the nanometer scale or below. Translating controlled operation of individual molecular machines to a larger dimension, for example, to 10-100 nm, which features many practical applications, is highly important but remains challenging. Here, we demonstrate a light-driven plasmonic nanosystem that can amplify the molecular motion of azobenzene through the host nanostructure and consequently translate it into reversible chiroptical function with large amplitude modulation. Light is exploited as both energy source and information probe. Our plasmonic nanosystem bears unique features of optical addressability, reversibility, and modulability, which are crucial for developing all-optical molecular devices with desired functionalities.**

When designing active nanoscale devices, three prerequisites are of paramount importance. First, an efficient energy source for triggering conformation changes at the nanoscale is crucial[1]. Equally important is the reversible control over conformation of individual nanostructures. Last but not least is the ability to report such nanoscale conformation changes and translate them into tunable functionalities[2,3]. Among a variety of energy sources, light represents a unique stimulus to power an operation[4]. Different from chemical fuels that unavoidably introduce contaminants in a system, light is clean and waste-free. Also, in contrast to chemical fuels that crucially depend on diffusion kinetics, light offers high spatial and temporal resolution as it can be switched on and off rapidly. Most importantly, light can deliver noninvasive readout of an optically active system, thus allowing for monitoring an operation in real time.

Here, we demonstrate an all-optically controlled plasmonic nanosystem in the visible range using DNA nanotechnology. Our system can amplify the sub-nanometer conformation changes of azobenzene through the active host nanostructure[5,6] and consequently translate the light-induced molecular motion of azobenzene into reversible plasmonic chiroptical response, which can be in-situ read out by optical spectroscopy. The plasmonic nanostructure comprises two gold nanorods assembled on a reconfigurable DNA origami template[7–14]. A photo-responsive active site is introduced on the template with an azobenzene-modified DNA segment[15]. Light can cyclically 'write' and 'erase' the conformation states of the nanostructure through photoisomerization of azobenzene at a localized region. Different conformation states are read by probe light.

**Results**

**Design of the photoresponsive nanostructures.** Photoisomerization of azobenzene[16] (see Figure 1) is widely used for construction of light-driven artificial molecular machines. In particular, azobenzene can be incorporated into DNA strands for reversible control of DNA hybridization[15,17–19](see Fig. 1b). Our active nanostructure is based on a three-dimensional (3D) reconfigurable DNA origami template (see Fig. 1c), which consists of two 14-helix bundles (80 nm × 16 nm × 8 nm), folded from a long single-stranded DNA scaffold with the help of hundreds of staple strands (see Supplementary Methods 1, Supplementary Tables 1,2 and Supplementary Figs. 1-5). The two linked origami bundles form a chiral object with a tunable angle[14] (see Supplementary Fig. 2). The active function of the structure is enabled by introducing an azobenzene-modified DNA segment on the template, which works as a recognition site to receive light stimuli. This photoresponsive segment comprises two DNA branches, which are extended from the two origami bundles, respectively. One branch possesses a double-stranded DNA (dsDNA) 20-base-pair part linked by disulfide bonds with azobenzene-modified oligonucleotides (Azo-ODN 1)[18,19]. The other branch contains Azo-ODN 2, which is pseudocomplementary to Azo-ODN 1. Azo-ODN 1 and Azo-ODN 2 contain three and four azobenzene modifications, respectively (see Supplementary Methods 1 and Supplementary Fig. 3). Multiple azobenzene modifications are essential for efficient photoregulation of DNA hybridization[17,20]. Upon UV light illumination, the azobenzene molecules in Azo-ODNs are converted to *cis*-form, resulting in dehybridization of the Azo-ODN duplex. The photoresponsive segment is opened and the conformation of the origami nanostructure is therefore 'relaxed'. In contrast, upon visible light illumination, the azobenzene molecules are converted to *trans*-form and Azo-ODNs can be hybridized into the Azo-ODN duplex. Therefore, the photoresponsive segment is locked. The dsDNA part is employed here in order to define a rigid angle between the two origami bundles for a stable chiral conformation.

**Photo-regulation of the DNA origami templates.** It has been reported that the illumination time and temperature affect the hybridization and dehybridization kinetics of the Azo-ODN duplex[18,20]. To ensure good switching efficiency and simultaneously avoid origami damage, the sample was kept at a temperature of 40 °C during all switching experiments[18]. As a representative case, the locked state of the origami template was with a right-handed conformation, in which the angle between the two bundles was approximately 50°. The sample was first illuminated by UV light (365 nm) for 15 min and then by visible light (450 nm) for 10 min.. Transmission electron microscopy (TEM) images of the sample after UV and visible light illumination are shown in Figs. 2a and 2c, respectively (for additional TEM images see Supplementary Figs. 5-11. Statistic histograms of the acute angle between two linked origami bundles based on an assessment of ~ 400 origami structures after UV and visible light illumination are presented in Figs. 2b and 2d, respectively and Supplementary Fig. 12. As shown in Fig. 2b, after UV light illumination, a broad distribution over angles is observed, with a maximum magnitude occurring around 90°. This reveals that the origami structures have been turned into the relaxed state by UV light. Here, 90° is more favorable owing to the electrostatic repulsion between the two bundles within one origami structure. On the other hand, after visible light illumination, a maximum magnitude over angles occurs around 50°, which is in accordance to our structure design (see Supplementary Fig. 2). This elucidates that a majority of the origami structures have been driven by visible light to the designated locked state. An enlarged-view of the origami structures in the locked state is shown in Fig. 2e. The dsDNA branch, which links the two origami bundles to define the angle, is clearly visible in the individual structures in Fig. 2e. An averaged TEM image reconstructed from the perfectly locked origami structures (~120) is presented in Fig. 2f. It demonstrates the excellent structural homogeneity and high angle accuracy within the locked structures. The TEM characterization reveals that *trans-cis* photoisomerization of azobenzene, which is associated with a molecular length change of ~ 3.5 Å[21] can be efficiently amplified by the origami structures into their distinct conformation changes on the order of 30 nm (see Supplementary Fig. 2). This corresponds to an amplification factor of ~ 100. Certainly, this amplification factor can be further increased by designing larger origami frames or larger angle changes. Moreover, given the remarkable precision of addressability afforded by DNA, this translation can be well controlled in an individual nanostructure at a localized region, which serves as an active recognition site in response to light stimuli.

**Light-driven 3D plasmonic nanosystem.** Positioning of plasmonic nanoparticles with high precision offered by DNA[22–30] further endows our light-driven systems with unique optical functionalities. To this end, two gold nanorods (AuNRs) are assembled on one origami template to form a 3D plasmonic chiral nanostructure (see Fig. 3a). Twelve binding sites on each origami bundle are extended with capture strands for robust assembly of one AuNR (38 nm×10 nm) functionalized with DNA complementary to the capture strands. The length of the binding site area is ~36 nm (see Supplementary Fig. 2). To ensure a high positioning accuracy of the AuNRs on origami, an additional thermal annealing procedure was carried out. Detailed information on the AuNR functionalization and assembly can be found in Supplementary Methods 2 and Supplementary Fig. 13..

When light interacts with the 3D chiral nanostructure, plasmons are excited in the two AuNRs that are placed in close proximity. The excited plasmons are collectively coupled in the cross conformation, leading to plasmonic chiroptical response[31–35]. The resulting plasmonic circular dichroism[36] (CD) spectra are very sensitive on conformation changes, ideal for optically monitoring the conformation evolution in real time[14].

Fig. 3b and Supplementary Fig. 14 show TEM images of the plasmonic nanostructures. A high assembly yield of the AuNR dimers on the origami templates has been achieved. Due to a higher affinity of the AuNRs to the carbon film of the TEM grid compared to that of DNA, the AuNR pairs appear side-by-side in the TEM images. To in-situ monitor the dynamic process associated with the conformation changes triggered by light, the CD response of a plasmonic sample was measured during visible (450 nm) and UV (365 nm) light illumination, respectively, using a J-815 CD spectrometer (Jasco). To be more specific, visible and UV light is used to 'write' and 'erase' the handed state of the plasmonic system, respectively, while circularly polarized light is used to 'read' the state changes. For a better elucidation, two representative CD spectra recorded after the system has achieved stable states for visible and UV light illumination are presented in Fig. 3c. The spectra were recorded within a wavelength range of 550 nm - 850 nm. The CD spectrum after visible light illumination is characterized by a bisignate dip-to-peak profile (in blue), which is typical for a right-handed system. This demonstrates that visible light has successfully driven the plasmonic system to the locked state, in which the conformation is 'written' as right-handed. On the other hand, the CD response after UV light illumination decreases significantly as shown by the purple curve. The plasmonic system has been converted into the relaxed state and the previous right-handed conformation is therefore 'erased'. A CD intensity modulation as high as 10 times between the two states has been achieved, demonstrating excellent photo-responsivity of the active plasmonic system. In this regard, molecular motion of azobenzene is spatially amplified by the host nanostructures and optically reflected through distinct chiroptical response changes. For the details of optical characterization, see Supplementary Methods 3 and Supplementary Figs. 15, 16.

To provide deeper insight, theoretical calculations were performed using the commercial software COMSOL Multiphysics based on a finite element method (see Supplementary Methods 4), and the results are shown in Supplementary Fig. 17. The CD spectra were calculated as difference of extinction for the left- and right-handed circularly polarized light. The assembled nanostructures were randomly dispersed in solution, and therefore averaging over different orientations was carried out. To account for the inhomogeneous spectral broadening resulting from the polydispersity of the AuNRs, the dielectric function of Au was modified by including an extra damping coefficient. Overall, the agreement between the experiment and theory is good.

Next, the CD intensities at 720 nm, *i.e*., approximately at the spectral dip position, are presented as a function of visible and UV light illumination time in Fig. 3d. The conversion from the right-handed state upon UV light illumination took approximately 15 min to achieve the stable relaxed state, whereas the conversion from the relaxed state upon visible light illumination took approximately 10 min to reach the stable right-handed state. The data curves can be well fit by first-order reaction kinetics with rate constants of $5\times10^{-3}$ s$^{-1}$ and $1.3\times10^{-2}$ s$^{-1}$ for UV and visible illumination, respectively. Also, the reversibility of the conversion is examined by alternative UV and visible light illumination in cycles for 15 and 10 min per exposure, respectively (see Fig. 4a and Supplementary Fig. 16). The CD intensity was recorded at 720 nm. As shown in Fig. 4b, excellent reversibility of the chiroptical response is achieved between the two states with large signal modulations. In brief, 'writing', 'erasing', and 'reading' actions can be coordinated efficiently with such a bi-stable system, in which each state can be converted into the other by light and consequently be reported by light.

## Discussion

The realization of light-driven plasmonic systems based on DNA nanotechnology offers many advantages for effectively manipulating materials and information at the nanoscale. From the material aspect, DNA as one of the most flexible materials in nanotechnology possesses unique biochemical specificity, remarkable spatial accuracy, and ease of addressability[37–39]. Inclusion of chemical species, which can execute reversible transformations by light such as azobenzene endows such hybrid systems with both spatial and temporal precision. From the information aspect, light as a stimulus renders 'writing' and 'erasing' of the conformation states in a reversible way possible without addition of any reagent. Meanwhile, light also serves as an information probe to read dynamic state changes in real time. Our plasmonic system may launch a new generation of sensing platforms[40,41], as light-induced structural changes could be optically tracked and successively retrieved. Finally, by harvesting light energy, individual nanostructures may generate collective actions and directly translate controlled molecular motion to a macroscopic level[42].

## Methods

**Materials.** DNA scaffold strands (p7650) were purchased from Tilibit Nanosystems. Unmodified staple strands (purification: desalting) were purchased from Eurofins MWG. Capture strands for the gold nanorods (AuNRs) (purification: desalting) were purchased from Sigma-Aldrich. Thiol-modified strands (purification: HPLC) were purchased from biomers.net. Azobenzene-modified DNA strands were obtained following a previously published procedure[18]. Agarose for electrophoresis and SYBR Gold nucleic acid stain were purchased from Life Technologies. Uranyl formate for negative TEM staining was purchased from Polysciences, Inc. AuNRs were purchased from Sigma-Aldrich (cat no. 716812). Other chemicals were purchased either from Carl-Roth or Sigma-Aldrich.

**Design and preparation of the DNA origami templates.** The design of the DNA origami structures was adopted from a previous study[14]. The strand routing diagram of the origami structures can be found in Supplementary Fig. 1. The sequences of the staple strands and modifications used for photoswitching are provided in Supplementary Table 1. The origami structures were prepared by thermal annealing (see Supplementary Table 2) and purified by agarose gel electrophoresis (see Supplementary Fig. 4) For details see Supplementary Methods 1.

**Light-driven conformational switching.** For UV illumination, a 3 Watt light emitting diode (Köhler Technologie-Systeme GmbH & Co. KG) with emission wavelength centered at 365 nm (±5 nm) was used. For visible light illumination, a white light emitting diode (M7RX, LED LENSER) and a bandpass optical filter centered at 450 nm with a 40 nm bandpass region (FB450-40, Thorlabs) were used. For TEM characterizations, the DNA origami sample was incubated at 40 °C and pH 8. First, the origami sample was exposed to UV light illumination for 15 min and part of the sample was used for TEM investigations. The rest of the sample was exposed to visible light illumination for 10 min and then used for subsequent TEM investigations.

**TEM characterization.** The DNA origami structures (with or without AuNRs) were imaged using a Philips CM 200 transmission electron microscope (TEM) operating at 200 kV. For imaging, the DNA origami structures (with or without AuNRs) were deposited on freshly glow discharged carbon/formvar TEM grids. The TEM grids were treated with a uranyl formate solution (0.75%) for negative staining of the DNA structures. Angles between the origami bundles in individual structures were obtained by manual analysis of the TEM images. The acute angles were chosen for analysis. Class average images (Fig. 2f) were obtained using EMAN2 software[43].

**Optical characterizations.** CD and UV-Vis measurements were performed with a J-815 Circular Dichroism Spectrometer (Jasco) using Quartz SUPRASIL cuvettes (105.203-QS, Hellma Analytics) with a path length of 10 mm. UV-Vis measurements were also performed with a BioSpectrometer (Eppendorf). Details can found in Supplementary Note 4.

**Acknowledgements**

We thank A. Jeltsch and R. Jurkowska for assistance with CD spectrometry. We thank M. Kelsch for assistance with TEM. TEM images were collected at the Stuttgart Center for Electron Microscopy (StEM). N. Liu was supported by the Sofja Kovalevskaja Award from the Alexander von Humboldt-Foundation. A. Kuzyk was supported by a postdoctoral fellowship from the Alexander von Humboldt-Foundation. A. Kuzyk and N. Liu were supported by a Marie Curie CIG Fellowship. We also thank for the financial support from the European Research Council (ERC) Starting Grant 'Dynamic Nano'. A. O. Govorov was supported by the U.S. Army Research Office under grant number W911NF-12-1-0407 and by Volkswagen Foundation (Germany). M. Endo was supported by JSPS KAKENHI (grant numbers 15H03837, 24104002, 26620133).

**Author Contributions:** A.K., and N.L. conceived the experiments. A.K. designed the DNA origami nanostructures. A.K., Y.Y., S.S., H.S and M. E. prepared the nanostructures. A.K. and S.S. performed TEM and CD characterization. X.D. carried out the theoretical calculations. A.O.G. offered useful suggestions. A.K. and N.L wrote the manuscript. All authors discussed the results, analyzed the data and commented on the manuscript.

**Additional information**

Supplementary information is available in the online version of the paper. Reprints and permissions information is available online at www.nature.com/reprints. Correspondence and requests for materials should be addressed to A.K or N.L.

**Competing financial interests**

The authors declare no competing financial interests.

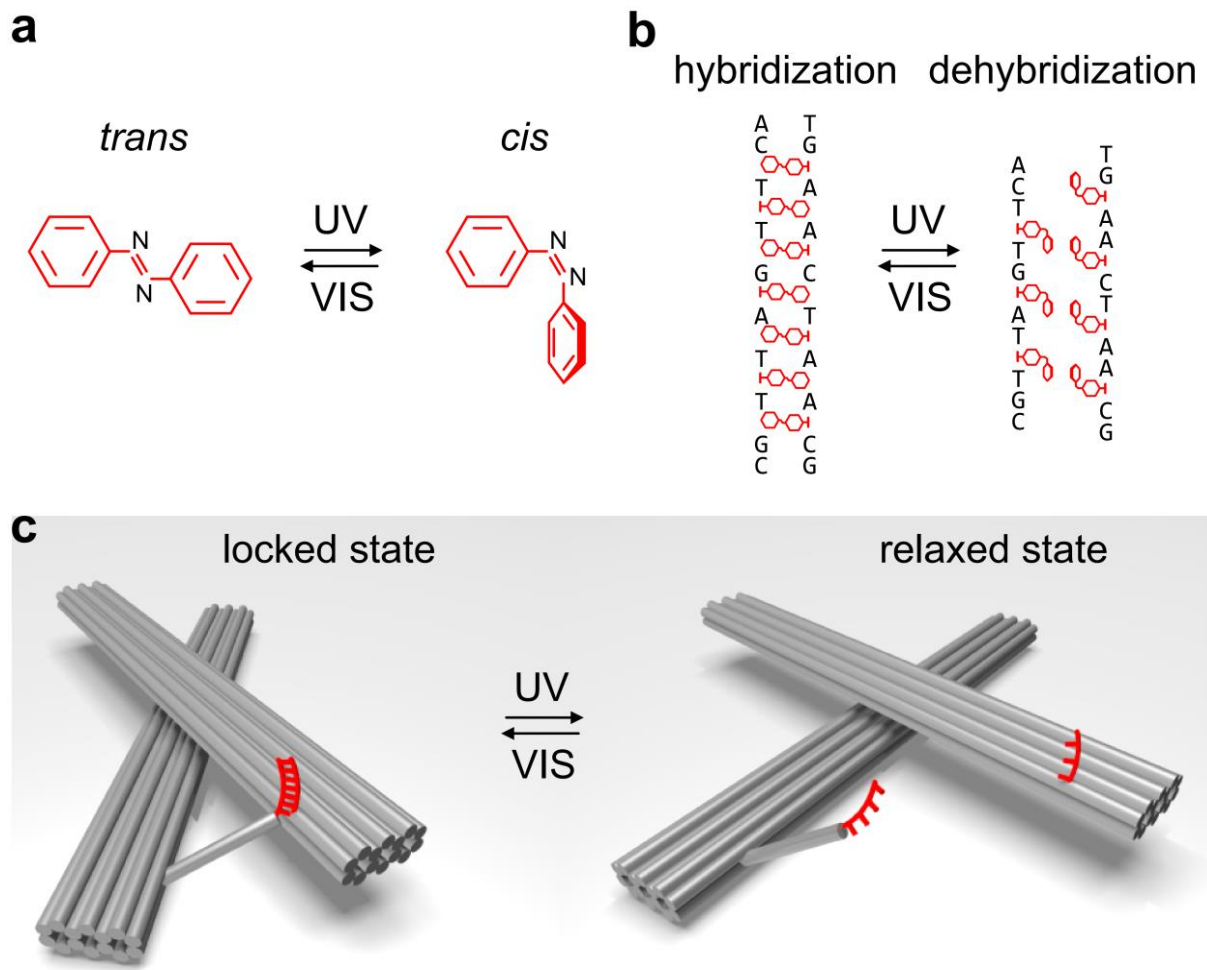

**Figure 1 | Light-induced conformation changes of DNA origami nanostructures. a,** *Trans-cis* photoisomerization of an azobenzene molecule by UV and visible (VIS) light illumination. **b,** Hybridization and dehybridization of azobenzene-modified DNA oligonucleotides controlled by *trans−cis* photoisomerization of azobenzene through UV and VIS light illumination. **c,** Photo-regulation of the DNA origami template between the locked and relaxed states by UV and VIS light illumination. The active function of the origami structure is enabled by introducing the azobenzene-modified DNA segment (red) in **b** on the template, which works as a recognition site to receive light stimuli for triggering light-induced motion.

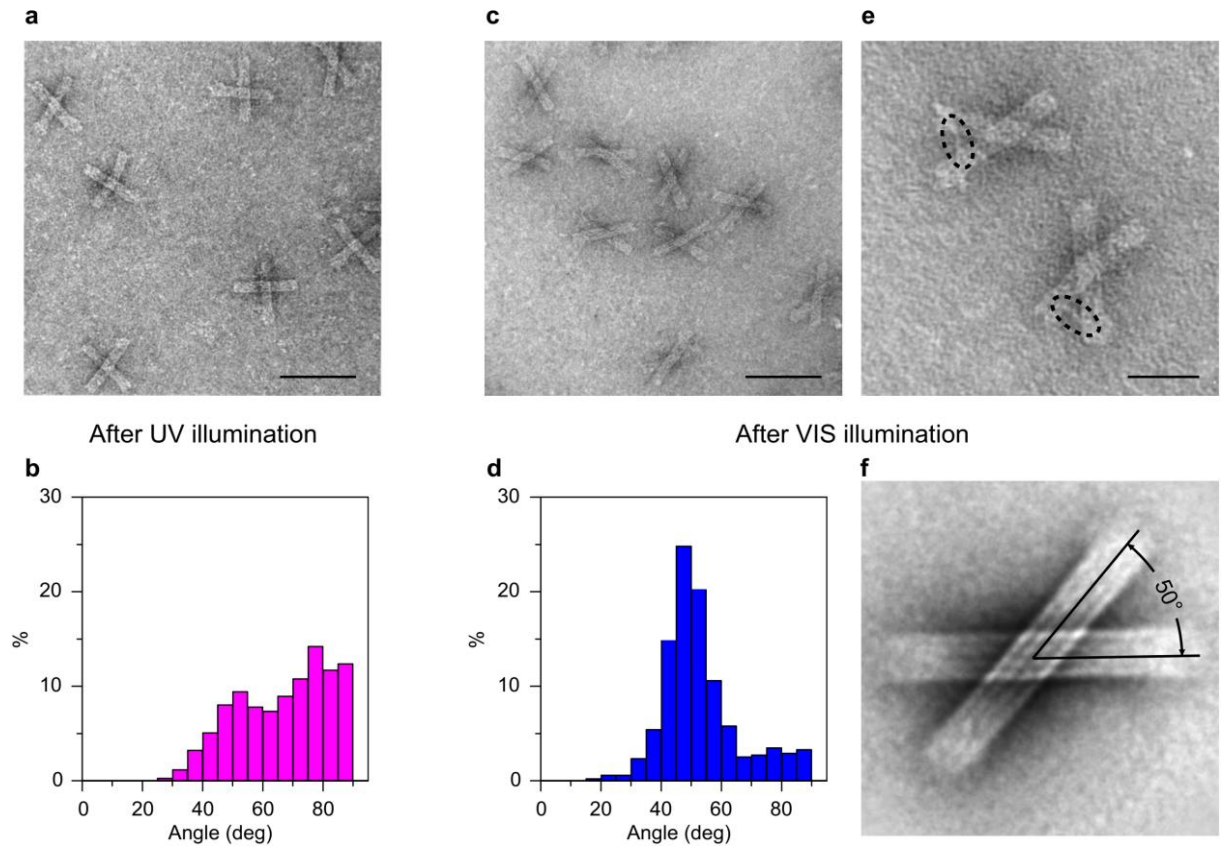

**Figure 2 | Structural characterization of the DNA origami nanostructures. a,** Transmission electron microscopy (TEM) image of the DNA origami nanostructures after UV light illumination **b,** Statistic histogram of the acute angle between two linked origami bundles after UV light illumination. The number of the analyzed structures: 463. A broad distribution over angles is observed. **c,** TEM image of the DNA origami nanostructures after VIS light illumination. The locked state is designed to be right-handed. **d,** Statistic histogram of the acute angle between two linked origami bundles after VIS light illumination. The number of the analyzed structures: 541. A maximum magnitude over angles occurs around 50°, which is in accordance to our structure design. **e,** Enlarged-view of the origami structures in the locked state. The dsDNA branch, which links the two origami bundles to define the angle, is clearly visible. **f,** Averaged TEM image reconstructed from locked origami structures. It evidently demonstrates the excellent structural homogeneity and high angle accuracy within the locked structures. Scale bars, 100 nm (**a,c**); 50 nm (**e**).

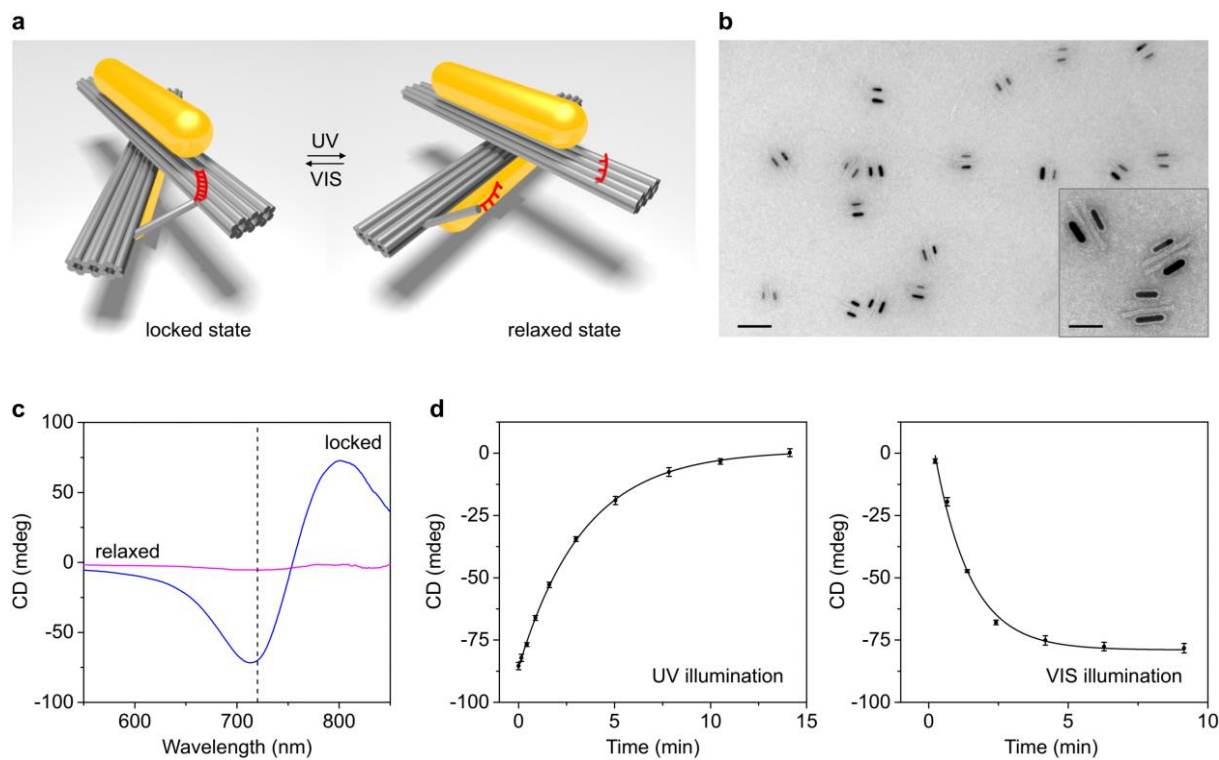

**Figure 3 | Light-driven 3D plasmonic nanosystem. a,** Schematic of the 3D plasmonic nanosystem regulated by UV and VIS light illumination for switching between the locked right-handed and relaxed states. Two gold nanorods are assembled on one origami template to form a 3D plasmonic chiral nanostructure. **b,** TEM images of the plasmonic nanostructures in the locked right-handed state. Scale bars are 200 nm and 50 nm in the large image and in the inset image, respectively. **c,** Measured circular dichroism (CD) spectra after UV (purple) and VIS (blue) illumination. **d,** Kinetic characterization of the 3D plasmonic nanostructures switching from the locked right-handed state to the relaxed state and vice versa upon UV and VIS illumination. The experimental data can be well fit by first-order reaction kinetics with rate constants of $5\times10^{-3}$ s$^{-1}$ and $1.3\times10^{-2}$ s$^{-1}$ for UV and VIS illumination, respectively.

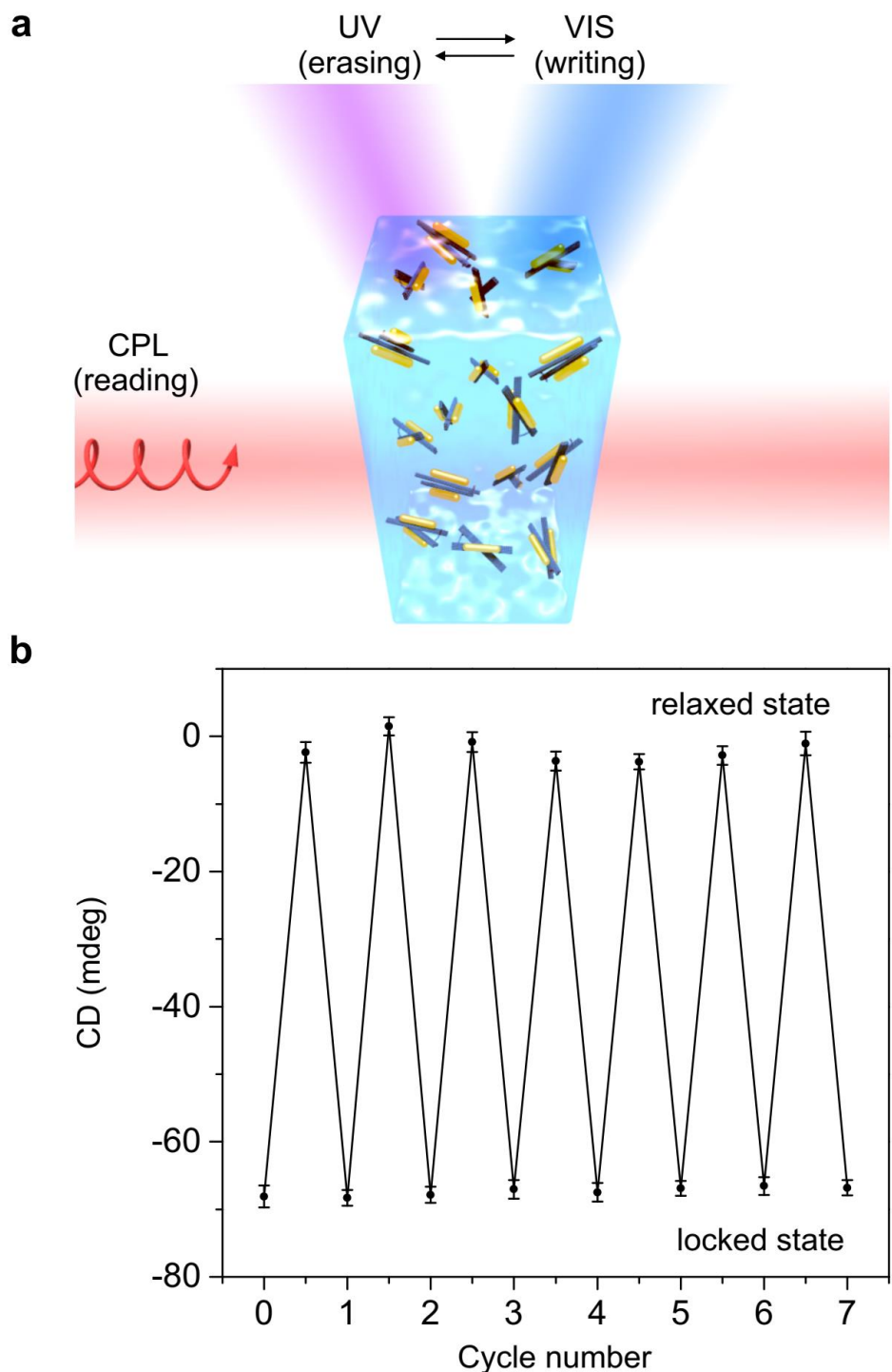

**Figure 4 | 'Writing', 'erasing', and 'reading' of the 3D plasmonic nanostructures by light. a,** Reversible conversion of the plasmonic nanostructures between the relaxed and locked right-handed states by UV and VIS light illumination, which performs the 'erasing' and 'writing' behavior, respectively. The resulting conformation states are probed by circularly polarized light (CPL) in real time, which performs the 'reading' behavior. **b,** CD intensity recorded at 720 nm (see Fig. 3c) during alternative UV and VIS illumination in multiple cycles. Excellent reversibility of the chiroptical response is achieved between the two states with large signal modulations.